\documentstyle[12pt,epsf]{article}
\baselineskip=\normalbaselineskip
\ifx\TwoupWrites\UnDeFiNeD\else\target{\magstepminus1}{11.3in}{8.27in}
	\source{\magstep0}{7.5in}{11.69in}\fi
\newfont{\fourteencp}{cmcsc10 scaled\magstep2}
\newfont{\titlefont}{cmbx10 scaled\magstep2}
\newfont{\authorfont}{cmcsc10 scaled\magstep1}
\newfont{\fourteenmib}{cmmib10 scaled\magstep2}
	\skewchar\fourteenmib='177
\newfont{\elevenmib}{cmmib10 scaled\magstephalf}
	\skewchar\elevenmib='177
\makeatletter
\newcommand\nonsequentialeqnum{
	\@addtoreset{equation}{section}
	\def\theequation{\arabic{section}.\arabic{equation}}}
\newif\ifp@bblock  \p@bblocktrue
\newcommand\nopubblock{\p@bblockfalse}
\newcommand\topspace{\hrule height 0pt depth 0pt \vskip}
\newcommand\p@bblock{\begingroup \tabskip=\hsize minus \hsize
	\baselineskip=1.5\ht\strutbox \topspace-2\baselineskip
	\halign to\hsize{\strut ##\hfil\tabskip=0pt\crcr
	\the\Pubnum\crcr
	\the\date\crcr}\endgroup}
\newcommand\YITPmark{\hbox{\fourteenmib YITP\hskip0.2cm
        \elevenmib Uji\hskip0.15cm Research\hskip0.15cm Center\hfill}}
\renewcommand\titlepage{\ifx\TwoupWrites\UnDeFiNeD\null\vspace{-1.7cm}\fi
	\YITPmark\vskip0.6cm
	\ifp@bblock\p@bblock \else\hrule height 0pt \relax \fi}
\makeatother
\newtoks\date
\newtoks\Pubnum
\newtoks\pubnum
\Pubnum={YITP/U-\the\pubnum}
\date={\today}
\newcommand{\frontpageskip}{\vspace{12pt plus .5fil minus 2pt}}
\renewcommand{\title}[1]{\frontpageskip
	\begin{center}{\titlefont #1}\end{center}\par}
\renewcommand{\author}[1]{\frontpageskip\par\begin{center}
	{\authorfont #1}\end{center}
	\nobreak
	}

\newcommand{\address}[1]{\par\begin{center}{\sl #1}\end{center}\par}

\renewcommand{\thanks}[1]{\footnote{#1}}
\renewcommand{\abstract}{\par\frontpageskip\centerline{\fourteencp Abstract}
	\vspace{8pt plus 3pt minus 3pt}}

\catcode`@=11
\@addtoreset{equation}{section}   
\def\theequation{\arabic{section}.\arabic{equation}}
\def\thebibliography#1{\section*{References\@mkboth
 {REFERENCES}{REFERENCES}}\list
 {\leftbibmark\arabic{enumi}\rightbibmark}{
 \settowidth\labelwidth{\leftbibmark #1\rightbibmark}\leftmargin\labelwidth
 \advance\leftmargin\labelsep
 \usecounter{enumi}}
 \def\newblock{\hskip .11em plus .33em minus -.07em}
 \sloppy\clubpenalty4000\widowpenalty4000
 \sfcode`\.=1000\relax}

\def\@cite#1#2{\leftcitemark{#1\if@tempswa , #2\fi}\rightcitemark}
\def\leftcitemark{[}
\def\rightcitemark{]}
\def\leftbibmark{[}
\def\rightbibmark{]}
\def\CITE#1{$^{\hbox{\small \cite{#1}}}$}

\catcode`@=10

\def\getlength#1{\ifx#1\end \let\next=\relax
    \else\advance\count0 by1 \let\next=\getlength \fi \next}
\def\emptybox#1#2{\framebox[#1]{\rule{0cm}{#2}}}   
\def\LeftFigure#1#2#3#4#5{
\begin{figure}
\noindent\emptybox{#1}{#2}\hfill
{\addtolength{\hsize}{-#1}\addtolength{\hsize}{-1cm}
\parbox[b]{\hsize}{\caption{\label{#3}#4}
{\count0=0 \getlength #5\end \ifnum\count0>0 \hfill\\ {\small #5} \fi}\par}
}
\end{figure}
}
\def\Figure#1#2#3#4#5{
\begin{figure}
\emptybox{#1}{#2}
\caption{\label{#3}#4} {\count0=0 \getlength #5\end \ifnum\count0>0 \hfill\\
{\small #5} \fi}
\end{figure}
}

\def\A{{\cal A}}

\def\F{{\cal F}}

\def\M{{\cal M}}

\def\T{{\cal T}}

\def\linebreak{\hfill\break}

\def\newterm#1{{\it #1}}
\def\Eq#1{Eq.(\ref{#1})}
\def\Eqs#1#2{Eqs.(\ref{#1})-(\ref{#2})}
\def\Prop#1{Proposition \ref{#1}}

\def\Theorem#1{Theorem \ref{#1}}

\def\tend{\rightarrow}

\def\equivalent{\quad\Leftrightarrow\quad}
\def\therefore{\mbox{\setbox0=\hbox{X}\hbox{$\ldotp$}\raise0.7\ht0\hbox{$\ldotp$}\hbox{$\ldotp$}} }
\def\because{\mbox{\setbox0=\hbox{X}\raise0.7\ht0\hbox{$\ldotp$}\hbox{$\ldotp$}\raise0.7\ht0\hbox{$\ldotp$}}\kern0pt }
\def\r#1{{\rm #1}}

\def\bm#1{\mbox{\boldmath $#1$}}

\def\Frac(#1/#2){\left(\frac{#1}{#2}\right)}
\def\Tr{\r{Tr}}

\def\Lie{\hbox{\rlap{$\cal L$}$-$}}
\def\vfield{\hbox{\rlap{$\cal X$}$-$}}

\def\RF{{\bm{R}}}

\def\orth{\perp}

\def\Hom(#1){{\rm Hom}(#1)}

\def\domain#1{{\rm dom}\; #1}

\def\support#1{{\rm supp}\;#1}

\def\maps{\rightarrow}
\def\mapsnamed#1{\stackrel{#1}{\longrightarrow}}

\def\nemapsnamed#1{\mbox{$\nearrow #1$}}
\def\semapsnamed#1{\mbox{$#1\searrow $}}

\def\In{\mathrel{\mbox{\setbox0=\hbox{$\cup$}\dimen0=\wd0\divide\dimen0 by 2
\box0\kern -\dimen0\vrule}}}
\def\Mapping#1#2#3#4{\begin{array}[t]{ccc}
#1 & \rightarrow & #3\\
\In &            & \In\\
#2 &             & #4 \end{array}}
\def\DiffG#1{\r{Diff}_0(#1)}
\def\SetDef#1#2{\left\{#1\;|\;#2\right\}}

\newtheorem{theorem}{Theorem}[section]
\newtheorem{proposition}{Proposition}[section]

\def\FilledSquare{\vbox{\setbox1=\hbox{X} \hrule height\ht1 width0.5\wd1
depth0pt}}
\newenvironment{proof}{\par\noindent {\bf Proof}\par}{\par\medskip}
\def\QED{ \FilledSquare}

\def\Beq{\begin{equation}}
\def\Eeq{\end{equation}}
\def\Beqr{\begin{eqnarray}}
\def\Eeqr{\end{eqnarray}}
\def\Beqrn{\begin{eqnarray*}}
\def\Eeqrn{\end{eqnarray*}}
\def\Bitm{\begin{itemize}}
\def\Eitm{\end{itemize}}

\begin{document}
\thispagestyle{empty}
%
\pubnum{95-18}
\date{July 1995}
\titlepage

\title{
Dynamics of Totally Constrained Systems \\
I. Classical Theory
}

\author{
Hideo Kodama
}
\address{
Yukawa Institute for Theoretical Physics, Kyoto University,
Uji 611, Japan
}

\abstract{
This is the first of a series of papers in which a new formulation of
quantum theory is developed for totally constrained systems, that is,
canonical systems in which the hamiltonian is written as a linear
combination of constraints $h_\alpha$ with arbitrary coefficients.
The main purpose of the present paper is to make clear that classical
dynamics of a totally constrained system is nothing but the foliation
of the constraint submanifold in phase space by the involutive system
of infinitesimal canonical transformations $Y_\alpha$ generated by the
constraint functions.  From this point of view it is shown that
statistical dynamics for an ensemble of a totally constrained system
can be formulated in terms of a relative distribution function without
gauge fixing or reduction. There the key role is played by the fact
that the canonical measure in phase space and the vector fields
$Y_\alpha$ induce natural conservative measures on acausal
submanifolds, which are submanifolds transversal to the dynamical
foliation. Further it is shown that the structure coefficients
$c^\gamma_{\alpha\beta}$ defined by
$\{h_\alpha,h_\beta\}=\sum_\gamma c^\gamma_{\alpha\beta}h_\gamma$
should weakly commute with $h_\alpha$,
$\sum_\gamma\{h_\gamma,c^\gamma_{\alpha\beta}\}\approx0$, in order
that the description in terms of the relative distribution function is
consistent. The overall picture on the classical dynamics given in
this paper provides the basic motivation for the quantum formulation
developed in the subsequent papers.
}

\newpage

\section{Introduction}

In the canonical approach to quantum gravity the Dirac quantization
prescription\CITE{Dirac.P1964B}
 is widely adopted, in which the classical constraints
$h_\alpha\approx0$ are formulated as constraints on physical states of
the form
\Beq
h_\alpha \Phi=0.
\label{DiracConstraint}\Eeq
As is well-know, however, this leads to the frozen formalism in which
the dynamical evolution equation is lost.  This problem is closely
related with the fact that operators corresponding to the physical
quantities which play the role of time are excluded from observables
in this formulation\CITE{Ashtekar.A&Stachel1991B}.
Further the Dirac quantization of general
relativity does not give a mathematically well-defined formulation
apart from the regularization and the operator ordering problems
because \Eq{DiracConstraint} does not have normalizable solutions in
general even if the spatial diffeomorphism freedom is eliminated
before quantization.

As discussed in Ref.\cite{Kodama.H1993}, this difficulty comes from
the too formal application of the Dirac procedure. Since the
hamiltonian is written as a linear combination of the constraint
functions in general relativity, the hamiltonian constraints carry all
the information on dynamics. Hence if we formulate the hamiltonian
constraint as the condition on state vectors as above, each state
vector becomes a dynamical object. This should be compared with the
ordinary quantum mechanics.  There state vectors are used to describe
dynamics of a system, but each state is not a dynamical
object. Dynamics is described by a one-parameter family of states
$\Phi(t)$ satisfying the Schr\"{o}dinger equation. Each state vector
in this family merely carries information by possible measurements at
each instant $t$.

This observation indicates that we should impose the hamiltonian
constraints on the object which picks up all the possible state
vectors allowed by dynamics. The most natural object of this nature
will be the probability amplitude $\Psi(\Phi)$ which assigns the
probability to each state vector, taking account of the probabilistic
nature of quantum mechanics.

Here one may notice that a similar phenomenon occurs in the classical
dynamics of a system with hamiltonian constraints. For simplicity let
us consider a system with a single hamiltonian constraint $h\approx0$
on a phase space $\Gamma$.  If we reduce this system to a canonical
system without constraint, dynamics is described by a curve
$\gamma_0(t)$ in a reduced phase space $\Gamma_0$ which is a solution
to a canonical equation of motion.  This curve corresponds to the
family of states $\Phi(t)$ in quantum mechanics.  On the other hand in
the original phase space this curve corresponds to a curve $\gamma$
contained in the constraint hypersurface $\Sigma_H$, which corresponds
to $\Psi$ above. This analogy becomes better if we consider an
ensemble of systems instead of a single system, for which
$\gamma_0(t)$ is replaced by a family of distribution functions
$\rho_0(t)$ on $\Gamma_0$ and $\gamma$ by a distribution function
$\rho$ on $\Gamma$. Clearly $\rho$ does not represent a state but is a
dynamical object which picks up possible states allowed by
dynamics. Hence $\rho$ should be constant along each curve in $\Gamma$
corresponding to a solution to the equation of motion in
$\Gamma_0$. This implies that it is not normalizable on $\Gamma$. This
may be regarded as the essential reason why the solutions to
\Eq{DiracConstraint} is unnormalizable.  It is obviously absurd to
postulate dynamics so that it picks up $\rho$ which is normalizable in
$\Gamma$. Applying the quantum hamiltonian constraints on states
just corresponds to such an approach.

This observation suggests that the investigation of the structure of
classical statistical dynamics of totally constrained systems will
shed a good light on how to find a consistent quantum formulation of
them.  This is the motivation of the present paper.  Since a variety
of forms exist for the canonical formulation of gravity and since the
structure of the problem is common in all the theories with general
covariance, we consider a generic totally constrained system in most
part of the paper.

The organization of the paper is as follows.  First in \S2 we consider
a simple totally constrained system obtained by embedding a canonical
system without constraint to a larger phase space in order to find how
to interpret the unnormalizable distribution function on the extended
phase space. Then in \S3 on the basis of the result obtained there we
describe how to formulate the statistical dynamics of a generic
totally constrained system with a single constraint without reduction
and not referring to special time variables. Further the general
structure of reduction and its freedom is examined because it is
relevant to the time variable problem.  In particular by applying it
to the totally constraint system describing a relativistic particle in
curved spacetime, it is shown that the background spacetime should
have a Killing vector in order that there is a natural reduction of
this system.  In the subsequent two sections the formulation obtained
for a single constraint system is extended to a multiple constraint
system. First in \S4 an overview on the canonical structure of general
relativity in terms of the ADM variables is given in order to make
clear that dynamics of a totally constrained system with multiple
constraints is completely determined by the foliation of the
constraint submanifold by the involutive system of the infinitesimal
canonical transformations generated by the constraint functions. Then
in \S5 based on this viewpoint, statistical dynamics for multiple
constraint systems is formulated in terms of the relative distribution
function by proving the existence of natural conservative induced
measures on acausal submanifolds. Section 6 is devoted to discussion.

\section{Embedding of an unconstrained system into a constrained system}

In this section we study the dynamics of a simple constrained system
obtained by embedding an unconstrained canonical system into a larger
phase space. The main purpose is to find the way how to formulate the
dynamics of a constrained system and its ensemble without reducing it
into an unconstrained one.

\subsection{Canonical system}

A canonical dynamical system with no constraint is specified by a
triplet $(\Gamma, \omega, h)$ of a phase space, a symplectic form and
a hamiltonian\CITE{Arnold.V1974B}.
The phase space $\Gamma$ is a $2n$-dimensional smooth
manifold, the symplectic form $\omega$ is a closed non-degenerate
2-form on $\Gamma$,
\Beqr
&& d\omega=0,\\
&& \Omega:={1\over n!}\wedge^n \omega={1\over n!}\omega^n \not=0,
\Eeqr
and the hamiltonian $h$ is a smooth function on $\RF\times\Gamma$.

Let $\F(\Gamma)$ and $\vfield(\Gamma)$ be the sets of all smooth
functions and all smooth vector fields on $\Gamma$, respectively.
Then for any $f\in\F(\Gamma)$, $\omega$ uniquely determines a vector
field $X_f\in\vfield(\Gamma)$, which is called
\newterm{the infinitesimal canonical transformation generated by $f$},
through the equation
\Beq
df = -I_{X_f}\omega  \quad \equivalent \quad
Vf=\omega(V,X_f) \quad \forall V\in\vfield(\Gamma),
\label{ICTrf:def}\Eeq
where $I_V$ is the inner product operator which maps a $p$-form $\chi$
to a $(p-1)$-form defined by
\Beq
(I_X \chi)(V_1, \cdots, V_{p-1})=\chi(X,V_1, \cdots, V_{p-1}).
\Eeq
 From the identity
\Beq
\Lie_X=I_X\circ d + d\circ I_X
\label{LieByIP&d}
\Eeq
the infinitesimal transformation $X_f$ preserves $\omega$ and $\Omega$;
\Beq
\Lie_{X_f}\omega=0,  \quad \Lie_{X_f}\Omega=0.
\Eeq
Conversely any vector field which satisfies this equation is an
infinitesimal canonical transformation generated by some function on
$\Gamma$, at least locally.

In terms of this infinitesimal canonical transformation the Poisson
bracket of two functions $f$ and $g$ are defined by
\Beq
\{f, g\} := -X_f g =- \omega(X_f, X_g).
\Eeq
It follows from this definition that
\Beq
X_{\{f,g\}}=-[X_f, X_g].
\Eeq
Thus the correspondence $f \mapsto X_f$ gives a homomorphism from
$\F(\Gamma)$ into $\vfield(\Gamma)$ as Lie algebras, whose kernel is
constant functions.

In appropriate local coordinates $(q^j, p_j)$ $\omega$ can be always
written as
\Beq
\omega = dp_j \wedge dq^j,
\Eeq
In this local coordinate system $X_f$ is expressed as
\Beq
X_f={\partial f\over\partial p_j}{\partial\over\partial{ q^j}}
-{\partial f\over\partial q^j}{\partial\over\partial{p_j}},
\Eeq
which leads to the familiar expression
\Beq
\{f,g\}={\partial f\over\partial q^j}{\partial g\over\partial p_j}
-{\partial f\over\partial p_j}{\partial g\over\partial q^j}.
\Eeq

Finally the dynamics in the phase space is determined by the
hamiltonian $h$ in the following way.  Let the canonical coordinate of
$\RF$ in $\RF\times\Gamma$ be $t$, and for each value of $t$ let
$Y:=X_h$ be the infinitesimal canonical transformation generated by
$h$ regarded as a function on $\Gamma$.  Then for a single system, its
possible histories are given by the integration curves of the vector
field $\partial_t+ Y$ on $\RF\times\Gamma$ when $t$ is regarded as the
time variable, and the value of $f\in\F(\Gamma)$ along each curve,
when regarded as a function of $t$, satisfies the equation
\Beq
{df\over dt}={\partial f\over\partial t}+\{f,h\}.
\Eeq
In particular in a local coordinate system $(t,u^a)$ of
$\RF\times\Gamma$ each integral curve follows the canonical equation
of motion
\Beq
{du^a\over dt}=Y^a=\{u^a,h\}.
\Eeq

On the other hand the behavior of an ensemble of the system is
described by a distribution function $\rho\in\F(\RF\times\Gamma)$,
which satisfies the two conditions: i)$\int_\Gamma \rho\Omega=1$ for
each $t$, and ii) $\rho\Omega$ is preserved by the equation of motion.
This second condition yields the equation of motion for $\rho$,
\Beq
\Lie_{\partial_t+Y}(\rho\Omega)=0 \equivalent (\partial_t+Y)\rho=0,
\label{EOM:DistributionFn:Reduced}\Eeq
which implies that $\rho$ is constant along each integration curves in
$\RF\times\Gamma$.  When we introduce $\rho_t$ defined by
$\rho_t(u)=\rho(t,u)(u\in\Gamma)$, the expectation value of
$f\in\F(\RF\times\Gamma)$ at a time $t$ is given by
\Beq
<f>_t = \int_\Gamma f\rho_t\Omega,
\Eeq
and its time-derivative by
\Beqr
&{\displaystyle{d\over dt}}<f>_t&=\int_\Gamma{\partial\over\partial
t}(f\rho_t)\Omega
=\int_\Gamma \left({df\over dt} \rho_t\Omega+ f\rho_t\Lie_Y\Omega\right)\\
&&=\int_\Gamma {df\over dt} \rho_t\Omega.
\Eeqr
In particular, when $f$ is a constant of motion, i.e., ${df\over
dt}=0$, $<f>_t$ becomes time-independent.

\subsection{Embedding into a totally constrained system}

The above canonical dynamical system can be embedded into totally
constrained canonical systems with larger phase spaces by various
ways. Here we consider the simplest one.

Let $(\tilde\Gamma,\tilde\omega, \tilde h)$ be a canonical system
defined by
\Beqr
&&\tilde\Gamma:=\RF^2\times\Gamma\ni (q^0,p_0,u),\\
&&\tilde\omega:=dp_0\wedge dq^0+ \omega,\\
&&\tilde h(q^0,p_0,u):=h(q^0,u)+p_0.
\Eeqr
Then the infinitesimal canonical transformation $\tilde Y$
generated by $\tilde h$ is expressed as
\Beq
\tilde Y=Y+\partial_{q^0} - \partial_{q^0}h \partial_{p_0}.
\label{Y:Reducible}\Eeq
Hence by the projection $\pi$ defined by
\Beq
\pi: \Mapping{\tilde\Gamma}{(q^0,p_0,u)}{\RF\times\Gamma}{(q^0,u)},
\Eeq
it is mapped to $\pi_*\tilde Y=Y+\partial_t$, and each integral curve
of $\tilde Y$ to a solution to the equation of motion in
$\RF\times\Gamma$.  Therefore, noting that $\tilde h$ is conserved and
$\pi$ is injective on each $\tilde h=$const surface, one sees that the
original canonical system is equivalent to the extended canonical
system with a constraint $\tilde h=$const.

In this embedding only the integral curves of $\tilde Y$ in the
extended phase space, which we call \newterm{the hamiltonian flow},
have a physical significance, while the canonical time variable for
the extended system, which is denoted by $\tau$, just plays the role
of a parameter of these curves. Hence for an arbitrary function
$N(\tau)$ the system with $\tilde h$ replaced by $H=N\tilde h$ is also
equivalent to the original system under the constraint $\tilde
h=$const.  In particular for the special choice of the constraint,
$\tilde h=0$, this equivalence holds for an arbitrary function $N \in
\F(\RF\times\tilde\Gamma)$ since
\Beq
X_{N\tilde h}= N \tilde Y + \tilde h X_N \approx N\tilde Y.
\Eeq
where $A \approx B$ means that $A=B$ under the constraint.  We express
this last situation by saying that the original canonical system is
embedded into \newterm{the totally constraint system} $(\tilde \Gamma,
\tilde \omega, \tilde h)$. In this expression $\tilde h$ is understood
to play the double roles, one as the generating function of the
hamiltonian flow and the other giving the constraint $\tilde h=0$.

\subsection{Distribution function on the extended phase space}

The distribution function $\rho$ for the unconstrained system, if it is
regarded as a function on $\tilde \Gamma$, is constant along the
hamiltonian flow from \Eq{EOM:DistributionFn:Reduced} and
\Eq{Y:Reducible}.  This is quite natural since each pure dynamical state of
the constrained system is not represented by a point but by a
hamiltonian flow line in the extended phase space.  Hence, taking
account of the constraint, it is natural to consider the distribution
function $\tilde \rho$ on the extended phase space defined by
\Beq
\tilde \rho = \rho \delta(\tilde h),
\Eeq
where $\delta(*)$ is the delta function. From this definition it
follows that $\tilde \rho$ is characterized as a distribution on
$\tilde\Gamma$ satisfying the two equations,
\Beqr
&& \tilde Y\tilde \rho=0,\\
&& \tilde h\tilde \rho=0.
\Eeqr
Here by a distribution on $\tilde\Gamma$ we mean a functional $f$ on
the space of smooth functions $\phi$ with compact supports in $\tilde
\Gamma$, which is expressed as
\Beq
<f,\phi>:=\int_{\tilde\Gamma}f\phi|\tilde\Omega|
\Eeq
when it can be identified with a function on $\tilde\Gamma$.

Since the original phase space $\Gamma$ at time $t$ can be identified
with the intersection of the $q^0=t$ hypersurface $\Sigma_t$ and the
constraint hypersurface $\Sigma_H$ in $\tilde\Gamma$, the expectation
value of $f\in\F(\RF\times\Gamma)$ at a time $t$ is expressed in terms
of $\tilde\rho$ and $E_t:=\delta(q^0 -t)$ as
\Beq
<f>_t = <E_t \tilde\rho, f>.
\label{ExpectationValue:Reducible}\Eeq
This fixes the interpretation and the normalization of the
distribution function $\tilde\rho$ in the extended phase space.

 From the dynamical point of view these $q^0=$const surfaces have no
special significance in the extended phase space, apart from that they
are `Cauchy surfaces' for $\tilde\rho$. In fact we can easily extend
the expression \ref{ExpectationValue:Reducible} to that for the
expectation value on an arbitrary hypersurface $\Sigma$ which is
transversal to all the hamiltonian flow lines.  Let us call such a
surface \newterm{a maximal acausal hypersurface} and denote the
corresponding expectation value by $<f>_\Sigma$.  Then for two maximal
acausal hypersurfaces $\Sigma_1$ and $\Sigma_2$, $<f>_{\Sigma_1}$ and
$<f>_{\Sigma_2}$ should coincide for any constant of motion $f$.  If
we express $<f>_\Sigma$ in terms of a measure $\mu_\Sigma$ as
\Beq
<f>_\Sigma=\int_\Sigma f d\mu_\Sigma,
\Eeq
the above condition implies that $\theta_*
\mu_{\Sigma_1}=\mu_{\Sigma_2}$ where $\theta$ is the diffeomorphism
from $\Sigma_1$ onto $\Sigma_2$ determined by the hamiltonian flow.
Hence the requirement that $\mu_{\Sigma_t}$ coincides with that given
by \Eq{ExpectationValue:Reducible} completely determines $\mu_\Sigma$
for any maximal acausal hypersurface.

The explicit form of the measure is given by
\Beq
d\mu_\Sigma=\tilde\rho|I_{\tilde Y}\tilde\Omega|_\Sigma,
\Eeq
where $|\chi|_\Sigma$ implies that the differential form $\chi$ is
regarded as a positive measure on $\Sigma$. In order to see this,
first note that, for any constant of motion $f$, $fI_{\tilde
Y}\tilde\Omega$
is a closed form from
\Beq
d(fI_{\tilde Y}\tilde\Omega)=\Lie_{\tilde Y}(f\tilde\Omega)-I_{\tilde
Y}d(f\tilde\Omega)=0.
\Eeq
Hence by applying the
Stokes' theorem to a region bounded by $\Sigma_1$ and $\Sigma_2$ one
gets $\int_{\Sigma_1}f d\mu_{\Sigma_1}=\int_{\Sigma_2}f
d\mu_{\Sigma_2}$.  On the other hand from the identity
\Beq
d\phi\wedge I_{\tilde Y}\tilde\Omega
= -I_{\tilde Y}(d\phi\wedge\tilde\Omega)+I_{\tilde Y}d\phi \tilde\Omega
=-\{\tilde h,\phi\}\tilde\Omega,
\Eeq
one can rewrite the above expression for $<f>_\Sigma$ as
\Beqr
&&<f>_\Sigma=<E_\Sigma\tilde\rho, f>; \nonumber\\
&&E_\Sigma=|\{\tilde h,\phi\}|\delta(\phi-\tau),
\label{ExpectationValue:Reducible1}\Eeqr
if the maximal acausal surface $\Sigma$ is specified by the condition
$\phi=\tau(=$const$)$. It is easy to see that the right-hand-side of
this equation coincides with \Eq{ExpectationValue:Reducible} for
$\phi=q^0$ and $\tau=t$.

\subsection{Probability interpretation of the relative distribution
function}

Since $\tilde\rho$ is a distribution on $\tilde\Gamma$, we are tempted
to interpret it simply as giving a probability measure for
measurements of physical quantities defined on $\tilde\Gamma$.
Following this interpretation, the probability for a set of quantities
$f_1,\cdots\in\F(\tilde\Gamma)$ to take value in a given set of ranges
$\Delta_1,\cdots \subseteq\RF$ should be given by
\Beq
\Pr(f_1\in\Delta_1,\cdots)=C <\tilde\rho, E_{f_1}(\Delta_1)\cdots>,
\Eeq
where $C$ is a normalization constant and $E_f(\Delta)$ is the
characteristic function of the region where the value of $f$ is
contained in $\Delta$.  However, this naive interpretation does not
work by itself because the integration of $\tilde \rho$ over the whole
extended phase space diverges:
\Beq
<\tilde \rho,1>=\int dt <1>_t =+\infty.
\Eeq
Hence $\tilde\rho$ cannot give a finite measure on $\tilde\Gamma$ by
simple renormalization. Nevertheless it can be interpreted as giving a
relative probability density under some limited situations.

To see this, let us define the conditional probability for
physical quantities $f_1,\cdots$ to take value in $\Delta_1,\cdots$
under the condition $\phi=\tau$, by
\Beq
\Pr(f_1\in\Delta_1,\cdots|\phi=\tau):=
\lim_{\begin{array}{c}|\Delta_\phi|\tend0\\ \tau\in\Delta_\phi\end{array}}
{\Pr(f_1\in\Delta_1,\cdots,\phi\in\Delta_\phi)\over
\Pr(\phi\in\Delta_\phi)}.
\label{InducedProbabilityMeasure:Simple}\Eeq
Then the expectation value of any $f\in\F(\tilde\Gamma)$ determined
from this probability coincides with the right-hand side of
\Eq{ExpectationValue:Reducible1}, if and only if $\{\tilde h,\phi\}$
is given by some function of $\phi$ on $\Sigma_H$. This condition
requires that $\phi$ is expressed in terms of a function $g$ and some
constant of motion $k$ as
\Beq
\phi \approx g(q^0+k).
\label{Condition:GoodTimeVariable}\Eeq
Let us call a function satisfying this condition \newterm{a good time
variable}.

When a maximal acausal hypersurface $\Sigma$ is given, we can always
find a good time variable which is constant on $\Sigma$. However, its
freedom is just a rescaling of the variable so that it is in general
impossible to find a good time variable which is constant on each of
more than two given acausal hypersurfaces.  Though this fact has no
significance in the classical framework, it seems to have a deep
implication in the quantum framework in connection with the unitarity
problem as will be discussed in the next paper\CITE{Kodama.H1995a}.

\section{Non-trivial system with a single hamiltonian constraint}

In this section we show that by a simple generalization of the
formulae in the previous section we can discuss the dynamics of a
totally constrained system with a single hamiltonian constraint
without reducing it to an unconstrained system.

\subsection{Dynamics in the extended phase space}

Let $(\Gamma, \omega, h)$ be a totally constrained system with a
single hamiltonian constraint $h=0$, and $Y\in\vfield(\Gamma)$ be the
infinitesimal canonical transformation $X_h$.  Then from the
consideration in the previous section, it is natural to interpret that
each integration curve of $Y$ on the constraint hypersurface
$\Sigma_H$ yields a possible time evolution of the system.  Hence if
$Y$ has a zero- point on $\Sigma_H$, it represents a solution for
which any physical quantity takes a fixed value.  Since such a
solution is quite unphysical, we assume that $Y$ does not vanish on
$\Sigma_H$.  On the other hand if there is a closed orbit in the
hamiltonian flow, it represents a completely periodic world like the
anti-de Sitter spacetime.  We do not consider such causality violating
cases in this paper either.  We further assume that the hamiltonian
flow is not ergodic. This is equivalent to requiring that the
foliation of $\Sigma_H$ by the hamiltonian flow has a locally trivial
bundle structure whose fiber is homeomorphic to $\RF$. Thus it has a
global section.  We call an extension of such a global section off
$\Sigma_H$ as hypersurface,
\newterm{a maximal acausal hypersurface}.

When we consider an ensemble of totally constrained systems with the
same structure, they are represented by a set of hamiltonian flow
lines in $\Gamma$, each of which intersects with a maximal acausal
hypersurface at a single point.  In the limit that the ensemble
consists of very large number of members, these intersection points
determine a measure $\mu_\Sigma$ on each maximal acausal hypersurface
$\Sigma$. From its definition this measure is preserved by the mapping
among maximal acausal hypersurfaces determined by the hamiltonian
flow.  Let $c$ be a positive definite constant of motion.  Then, since
$cI_Y\Omega$ is a closed form and yields a measure with the same
property on each maximal acausal hypersurface as shown in the previous
section, the Radon-Nykodim derivative of $\mu_\Sigma$ by
$c|I_Y\Omega|$ yields a function which is constant along the
hamiltonian flow on the constraint surface $\Sigma_H$.  Thus if we
regard this function as a distribution whose support is contained in
$\Sigma_H$, we are naturally led to the distribution function $\rho$
on $\Gamma$ which satisfies
\Beqr
&& Y\rho=0,\label{Eq:DistributionFunction:Single1}\\
&& h\rho=0 \label{Eq:DistributionFunction:Single2}
\Eeqr
in the distribution sense. From its definition the expectation value
of a physical quantity $f\in\F(\Gamma)$ on a maximal acausal
hypersurface $\Sigma$ for the ensemble is given by
\Beq
<f>_\Sigma=\int_\Sigma fc\rho|I_Y\Omega| / \int_\Sigma c\rho|I_Y\Omega|.
\Eeq

In practical situations each maximal acausal hypersurface is specified
by the condition $\phi=$const in terms of a physical quantity $\phi$
on the phase space $\Gamma$. Here the constant should be a special
value and the other values may not give maximal acausal hypersurface
in general.  We call such a function \newterm{instant function}.

In terms of the instant function the dynamics of the totally
constrained system is formulated in the following way. First one
selects an appropriate instant function $\phi_1$ from measured
quantities, and specifies a maximal acausal hypersurface, say, by
$\phi_1=0$.  Then measurements of various physical quantities
determines a measure $\mu_{\Sigma_1}$, which in turn determines the
value of the distribution function $\rho$ on the maximal acausal
hypersurface $\Sigma_1$. $\rho$ is uniquely extended over the phase
space by
\Eqs{Eq:DistributionFunction:Single1}{Eq:DistributionFunction:Single2},
at least around the constraint hypersurface $\Sigma_H$. Once the
distribution function is determined, one can calculate the expectation
value of any physical quantity at an instant specified by any instant
function.  Here, though the distribution function depends on the
choice of the constant of motion $c$ in the above procedure, this
freedom does not affect the predictions on the expectation values
since $c$ and $\rho$ come into the theory always in the combination
$c\rho$.

The reason why we have introduced the apparently superfluous freedom
of $c$ in the definition of $\rho$ is to widen the concept of good
time variables introduced in the previous section.  Let $\phi$ be an
instant function such that $\phi=\tau(=$const) gives a maximally
acausal hypersurface for any $\tau$ in some open interval of $\RF$.
If we require that the probability measures on these hypersurfaces
derived from the natural measure $\rho\Omega$ as in
\Eq{InducedProbabilityMeasure:Simple} coincides with the conserved
measure $\mu_\Sigma$, we obtain the condition
\Beq
c\{h,\phi\}\approx f(\phi),
\Eeq
where $f(\phi)$ is an appropriate function of $\phi$.  This condition
is equivalent to the condition that $\phi$ is a function of $\phi_0$
which is a solution to the equation
\Beq
cY\phi_0\approx \r{const}.
\label{Single:GoodTimeVariable}\Eeq
We will show later that we must allow for a nontrivial choice of $c$
for a natural time variable to satisfy this condition even in simple
cases.

\subsection{Reduction}

As clarified in the previous section, the dynamics of a totally
constrained system $(\Gamma,\omega,h)$ can be described with no
reference to a special time variable.  Now let us study the relation
of this description with that in terms of a time variable in a
canonical system with no constraint which is obtained by reduction.

In general a canonical dynamical system without constraint
$(\Gamma_0,\omega_0,h_0)$ is a reduction of the totally constrained
system $(\Gamma,\omega,h)$ if there exists a diffeomorphic embedding
\Beq
\Phi: \RF\times\Gamma_0 \maps \Sigma_H
\Eeq
which satisfies the following two conditions:
\begin{itemize}
\item[i)] $\Phi_*(\partial_t +  Y_0)=k Y \quad(k\in\F(\Gamma))$,
\item[ii)] $\Phi^*\omega(Z_1,Z_2)=\omega_0(Z_1,Z_2) \quad
Z_1,Z_2 \in\vfield({t}\times\Gamma_0)$,
\end{itemize}
where $Y_0$ is the infinitesimal canonical transformation on
$\Gamma_0$ generated by $h_0$.  A convenient characterization of
$\Phi$ is given by the following well-known result.

\begin{proposition}\label{Prop:ReductionMapping}
The necessary and sufficient condition for the mapping $\Phi$ to give
a reduction is that the following equation holds:
\Beq
\Phi^*\omega=\omega_0 -dh_0\wedge dt.
\Eeq
\end{proposition}
\begin{proof}
\noindent
1) Necessity.

 From the condition ii) for the reduction mapping there exists a 1-form
$\xi$ such that $\Phi^*\omega=\omega_0-\xi\wedge dt$.  Since $\omega$
and $\omega_0$ are both closed, we obtain $d\xi\wedge dt=0$.  From
this it follows that we can choose $\xi$ so that it is closed. Hence
$\Phi^*\omega$ can be written as $\Phi^*\omega=\omega_0-dp\wedge dt$
in terms of some function $p$. Applying $I_{\partial_t+Y_0}$ on this
expression leads to
$$
I_{\partial_t+Y_0}\Phi^*\omega=d(p-h_0)-(\partial_t p+Y_0 p)dt.
$$
On the other hand from the condition ii)
$$
I_{\partial_t+Y_0}\Phi^*\omega=\Phi^*(kI_Y\omega)=-\Phi^*(kdh)=0.
$$
Hence we obtain $dp=dh_0+(\partial_t p + Y_0 p)dt$, which leads to the
equation in the proposition.

\noindent
2) Sufficiency.

If the equation in the proposition holds, the condition ii) is
obvious.  Further it is easy to check that
$\Phi^*I_{\Phi_*(\partial_t+Y_0)}\omega
=I_{\partial_t+Y_0}\Phi^*\omega=0$. Hence there is a function $k$ on
$\Gamma$ such that
$$
I_{\Phi_*(\partial_t+Y_0)}\omega\approx -kdh=I_{kY}\omega.
$$
This is equivalent to the condition i). \QED
\end{proof}

The reduction mapping $\Phi$ induces a function $\phi$ and a vector
field $V$ on $\Sigma_H$ defined by
\Beq
t=\Phi^*\phi, \quad \Phi_*\partial_t =V,
\Eeq
which are related by
\Beq
V\phi=1.
\label{ReductionField:Normalization}\Eeq
The function $\phi$, when extended off the constraint surface
$\Sigma_H$, yields a time variable on $\Gamma$, and foliates $\Gamma$
into a family of acausal hypersurfaces
$\Sigma_t=\{u\in\Gamma|\phi(u)=t\}$.  On the other hand the vector
field $V$, when extended off $\Sigma_H$, generates a one-parameter
family of transformations $\nu_\tau$ on $\Gamma$ such that
$\nu_\tau(\Sigma_t)=\Sigma_{t+\tau}$, and $\F(\Gamma_0)$ can be
identified with the restriction to $\Sigma_H$ of the set of functions
on $\Gamma$ that are invariant by these transformations.  Hence the
reduced phase space $\Gamma_0$ can be naturally identified with
$\Sigma_0\cup\Sigma_H$. Under this identification the reduction
mapping $\Phi$ can be written as
\Beq
\Phi:\Mapping{\RF\times\Gamma_0}{(t,x)}{\Sigma_H}{\nu_t(x)}.
\Eeq
Hence, when an acausal hypersurface $\Sigma_0$ is given, the vector
field $V$ completely determines the reduction.  For this reason we
call $V$ \newterm{the reduction field}.

The reduction field yields a reference dynamics in describing the
dynamics of the totally constrained system by time evolution, and the
hamiltonian in the reduced system is essentially the generator of the
deviation of the hamiltonian flow from this reference dynamics.  To
see this, let $\theta_t$ be the mapping from $\Sigma_0$ to $\Sigma_t$
determined by the hamiltonian flow. Then $\eta_t=\nu_{-t}\theta_t$
yields a family of transformations on $\Gamma_0=\Sigma_0\cap\Sigma_H$
describing the deviation of the hamiltonian flow and the reference
dynamics.  For a given point $u\in\Gamma$ the tangent vector
$Z_{\eta_t(u)}(t)$ of the curve $\eta_t(u)$ is given by
\Beq
Z_{\eta_t(u)}(t)=-V_{\eta_t(u)}+(\nu_{-t})_*(kY)_{\theta_t(u)}
=(\nu_{-t})_*(-V+kY)_{\theta_t(u)},
\Eeq
where $k$ is a function determined by the condition $kY\phi\approx1$.
Hence from the condition ii) for the reduction mapping we see that
$Z(t)$ is nothing but the infinitesimal canonical transformation $Y_0$
generated by $h_0$.

An arbitrary vector cannot be a reduction field. It must be
approximately a canonical transformation as the following proposition
shows.

\begin{proposition}\label{Prop:ReductionTheorem}
A vector field $V$ on $\Gamma$ is a reduction field if and only if
it satisfies the following condition:
\begin{itemize}
\item[i)] $Vh\approx0$,
\item[ii)] there exists a function $\phi$ on $\Gamma$ such that
$V\phi\approx1$ and $\Lie_V\omega\wedge d\phi|_{\Sigma_H}=0$.
\end{itemize}
In particular any infinitesimal canonical transformation satisfying i)
is a reduction field. For this case the reduced hamiltonian $h_0$ can
be chosen to be independent of $t$.
\end{proposition}

\begin{proof}
\noindent
1) Necessity

 From the definition of the reduction field i) is obvious, and for the
time variable $\phi$ in the reduction the former half of condition ii)
is satisfied by definition. Further since $\Phi^*\omega=
\omega_0-dh_0\wedge dt$ from \Prop{Prop:ReductionMapping},
$$
\Phi^*(dI_V\omega\wedge d\phi)=dI_{\partial_t}\Phi^*\omega\wedge dt=0.
$$
This is equivalent to the latter half of ii).

\noindent
2) Sufficiency

By taking the function $\phi$ in condition ii) as a time variable, let
us construct the mapping $\Phi: \RF\times\Gamma_0 \maps \Sigma_H$ from
$\phi$ and $V$ exactly in the same way as described below
\Eq{ReductionField:Normalization}. Then it obviously follows
that $\Phi_* \partial_t=V$ and $\Phi^*\phi=t$. Hence from ii) we
obtain $0=\Phi^*(dI_V\omega\wedge d\phi)=d(I_{\partial_t}\Phi^*\omega)
\wedge dt$. This implies that there exist $h_0,\alpha\in\F(\RF\times\Gamma_0)$
such that  $I_{\partial_t}\Phi^*\omega=dh_0+\alpha dt$. Since $I_X^2=0$,
$\alpha$ is expressed as $\alpha=-\partial_t h_0$. Hence $\Phi^*\omega$
is written in terms of 2-form $\omega_0$ such that $I_{\partial_t}\omega_0=0$
as
$$
\Phi^*\omega=\omega_0-dh_0\wedge dt.
$$
Since $\omega$ is closed, $\omega_0$ is also closed. Further from
$$
\Lie_{\partial_t}\omega_0\wedge dt=\Lie_{\partial_t}\Phi^*\omega\wedge dt
=\Phi^*(\Lie_V\omega\wedge d\phi)=0,
$$
we obtain $\Lie_{\partial_t}\omega_0=0$. Hence from
\Prop{Prop:ReductionMapping} $\Phi$ yields a reduction mapping.\QED
\end{proof}

Though the reduction fields are rather restricted, we can find a
reduction $\Phi$ for which $h_0=\Phi^* p$, for an arbitrarily given
function $p$ on $\Gamma$. To see this, let $\phi$ be a function on
$\Gamma$ such that $Y\phi=-\{h,\phi\}\not=0$ on $\Sigma_H$, and put
$V=X_p+\alpha X_\phi$ where $p$ and $\alpha$ are functions to be
determined so that $V$ is a reduction field. First from condition i)
in
\Prop{Prop:ReductionTheorem} $\alpha$ is uniquely determined as
$\alpha=-\{h,p\}/\{h,\phi\}$. Since $\Lie_V\omega=-d\alpha\wedge
d\phi$, condition ii) simply reduces to $V\phi=-X_\phi p\approx1$. As
$X_\phi$ is transversal to $\Sigma_H$, we can always find a solution
$p$ to this equation such that $p$ coincides with an arbitrarily given
function on $\Sigma_H$. Since $h_0=\Phi^*p|_{\Sigma_H}$ and $Vp\approx
Yp/Y\phi$, this implies that $h_0$ can be an arbitrary function on
$\RF\times\Gamma_0$.

This observation shows that reduction and the corresponding reduced
hamiltonian has a physical significance only when the original system
has some kind of time-translation symmetry which induces a reduction
field in the phase space. In generic cases for which no such symmetry
exits it is more natural to discuss the dynamics in the original phase
space with a constraint as described in the previous section.

\subsection{Example: a relativistic particle in curved background}

We illustrate the argument so far by a simple totally constrained
system describing a relativistic particle moving on a curved
background $(M,g)$ with a mass which may be position-dependent.

The action for this system is given by
\Beq
S_2=-2\int d\tau\left[-U(x)g_{\mu\nu}(x)\dot x^\mu\dot x^\nu\right]^{1/2}.
\Eeq
This action is equivalent to
\Beq
S_1=\int d\tau\left\{p_\mu(\dot x^\mu-v^\mu)
-2\left[-U g_{\mu\nu}v^\mu v^\nu\right]^{1/2}\right\},
\Eeq
where $p_\mu$ is a Lagrange multiplier.  The variation of this action
with respect to $v^\mu$ yields
\Beq
p_\mu={1\over N}g_{\mu\nu}v^\nu; \quad
N:= {1\over 2U}\left[-U g_{\mu\nu}v^\mu v^\nu\right]^{1/2}.
\Eeq
By eliminating $v^\mu$ in $S_1$ with the help of this equation we get
\Beq
S=\int d\tau\left[p_\mu\dot x^\mu
- {1\over2}N(g^{\mu\nu}p_\mu p_\nu +U) \right],
\Eeq
where $N$ is regarded as an independent variable.  Hence the original
system is equivalent to the totally constrained canonical system
\Beqr
&(\Gamma,\omega,h): &  \Gamma=T^* M,\nonumber\\
&		    & \omega=d\theta=dp_\mu\wedge dx^\mu,\nonumber\\
&		    & h={1\over 2}\left[g^{\mu\nu}(x)p_\mu p_\nu +
U(x)\right],\label{Constraint:ParticleInCurvedST}
\Eeqr
where $T^*M$ is the cotangent bundle of $M$ and $\theta$ is its canonical
1-form.

This totally constrained system is not simply reducible as the
hamiltonian constraint is quadratic in the momentum unlike the system
considered in \S2.  However, one can discuss its dynamics in the
sense discussed in this section except for special cases.  In fact,
the generating vector $Y$ of the hamiltonian flow of this system is
given by
\Beq
Y=p^\mu{\partial\over\partial x^\mu}
-{1\over2}\left({\partial g^{\lambda\sigma}\over \partial x^\mu}p_\lambda
p_\sigma
+{\partial U\over \partial x^\mu}\right){\partial\over\partial p_\mu},
\Eeq
which vanishes at points where $p_\mu=0$ and $\partial U/\partial
x^\mu=0$.  Hence if there is no point such that $U=0$ and
$\partial U/\partial x^\mu=0$, the hamiltonian flow has global acausal
hypersurfaces.

As shown in the previous section, we can always find a reduction of
this system into an unconstrained system.  However, in order for the
corresponding reduction field to be associated with some time
translation of the system, the system must have a symmetry.  To see
this, let $K$ be a vector field on $M$. Then it induces a vector field
$\tilde K$ on the phase space $T^*M$ which is expressed in the local
coordinate system $(x^\mu,p_\nu)$ as
\Beq
\tilde K=K^\mu{\partial \over \partial x^\mu}
- p_\nu \partial_\mu K^\nu {\partial\over \partial p_\mu}.
\Eeq
It is easy to see that this is an infinitesimal canonical
transformation generated by the function $p_\mu K^\mu$. The condition
for this field to be a reduction field is given by the following
proposition.

\begin{proposition}\label{Prop:KillingReduction}
The vector field $\tilde K$ yields a reduction field of a totally
constrained system $(T^*M,\omega,h)$ if and only if $\tilde
Kh\approx0$. In particular for the constraint $h$ given by
\Eq{Constraint:ParticleInCurvedST} this is equivalent to the condition
that $K$ is a Killing vector of the metric $\tilde g=U g$.
\end{proposition}
\begin{proof}
Since $\tilde K$ is an infinitesimal canonical transformation the
former half of the proposition is obvious from
\Prop{Prop:ReductionTheorem}.  For $h$ given by
\Eq{Constraint:ParticleInCurvedST} the condition is expressed as
$$
0\approx \tilde K(U^{-1}h)=-\tilde\nabla_\mu(\tilde g_{\nu\alpha} K^\alpha)
\tilde g^{\mu\lambda}\tilde g^{\nu\sigma}p_\lambda p_\sigma,
$$
where $\tilde \nabla_\mu$ is a covariant derivative with respect to
$\tilde g$.  This equation is equivalent to
$$
\tilde\nabla_\mu(\tilde g_{\nu\alpha} K^\alpha)
+\tilde\nabla_\nu(\tilde g_{\mu\alpha} K^\alpha)=0,
$$
which implies that $K$ is a Killing vector of the metric $\tilde g$.\QED
\end{proof}

This restriction on the system is closely related with the condition
for the system to have a good time variable which is independent of
$p_\mu$.  To see this, let $\phi$ be a function on $M$. Then, since
$Y\phi$ is now given by
\Beq
Y\phi=g^{\mu\nu}p_\nu{\partial\phi\over\partial x^\mu},
\Eeq
the condition for $\phi$ to be a good time variable,
\Eq{Single:GoodTimeVariable}, is written as
\Beq
p^\mu p^\nu \nabla_\mu \nabla_\nu \phi
-{1\over2}\nabla^\mu\phi\nabla_\mu U \approx0,
\Eeq
which is equivalent to
\Beq
U\nabla_\mu \nabla_\nu \phi
=-{1\over2}g_{\mu\nu}\nabla^\lambda\phi\nabla_\lambda U.
\Eeq
This last equation is written in terms of the covariant derivative
with respect to the metric $\tilde g_{\mu\nu}=U g_{\mu\nu}$ as
\Beq
\tilde\nabla_\mu(U\tilde\nabla_\nu\phi)
+\tilde\nabla_\nu(U\tilde\nabla_\mu\phi)=0.
\Eeq
Hence the metric $\tilde g_{\mu\nu}$ must have a Killing vector $K$
such that
\Beq
K^\mu= \nabla^\mu\phi.
\Eeq
In particular $\tilde g_{\mu\nu}$ must be static.

Putting these two requirements together, we find that there should
exists a function $\phi$ such that $\nabla^\mu\phi$ is a Killing
vector of $\tilde g$ and $\nabla_\mu\phi\nabla^\mu\phi=$const in order
that there is a good time variable which is a function on $M$ and
whose gradient field generates a reduction field. It is easy to see
that these conditions are satisfied if and only if $g$ has a static
Killing vector along which $U$ is constant(cf. Kucha\v{r}'s article in
Ref.\cite{Ashtekar.A&Stachel1991B}. For such cases the
reduced hamiltonian is given by $p_\mu\nabla^\mu \phi$.

For example for a relativistic free particle in Minkowski spacetime
for which $U=m^2=$const, the translation Killing vector
$a^\mu\partial_\mu$ satisfies these conditions when $a^\mu$ is a
constant time-like vector.  The time function and the reduced
hamiltonian are given by $\phi=a_\mu x^\mu$ and $h_0=a^\mu p_\mu$. On
the other hand the boost and rotation Killing vectors do not
correspond to good time variables on $M$ because they are not gradient
vectors.

\section{General relativity as a totally constrained system}

Before extending the argument on the totally constrained system with a
single constraint to a more generic case, we give an overview on the structure
of the totally constrained system obtained from general relativity
\CITE{Fischer.A&Marsden1979B}.
The main purpose is to make clear that the classical dynamics of
general relativity as a totally constrained system is nothing but a
foliation of the constraint submanifold such that each leaf is
one-to-one correspondence with a 4-dimensional diffeomorphism class of
solutions to the Einstein equations.  This fact will be used to
establish an interpretation of generic systems in the next section.

\subsection{ADM canonical formulation on the 3-metric space}

For simplicity we only consider globally hyperbolic vacuum spacetime
$(M,g)$, and assume that it is spatially compact. Hence $M$ is
diffeomorphic to $\RF\times S$ where $S$ is a compact space.  Let
$q(t)$ be the induced 3-metric on $\{t\}\times S$, $K(t)$ the extrinsic
curvature tensor of $\{t\}\times S$, and $n(t)={1\over N}
(\partial_t-\nu)$ the unit normal to $\{t\}\times S$ where
$\nu\in\vfield(\{t\}\times S)$. Then by regarding $q(t)$, $K(t)$,
$N(t)$ and $\nu(t)$ as quantities on $S$ with the time parameter $t$,
$K(t)$ is written as
\Beq
K_{jk}={1\over 2N}(-\partial_t q_{jk}+ D_j\nu_k + D_k\nu_j),
\Eeq
where $D_j$ is the covariant derivative with respect to $q$. Further
the Einstein-Hilbert action for the spacetime $(M,g)$,
\Beq
S_2={1\over 2\kappa^2}\int_M d^4x\,\sqrt{-|g|}R,
\Eeq
is expressed in terms of these quantities as
\Beq
S_2={1\over 2\kappa^2}\int dt\int_S d^3x \sqrt{|q|}N({}^3\!R + q(K,K)-(\Tr
K)^2).
\Eeq
By introducing the momentum variable $p^{jk}$ conjugate to $q_{jk}$ by
\Beq
p^{jk}=-{\sqrt{|q|}\over2\kappa^2}(K^{jk}-q^{jk}\Tr K),
\Eeq
This action is put into the canonical form,
\Beq
S=\int dt\left[<p,\dot q>-\left(<h_\orth,N>+<h_D,\nu>\right)\right],
\label{GR:CanonicalAction}\Eeq
where $h_\orth$, $h_D$ and $p$ are linear functionals of functions, vector
fields and 2nd-rank covariant tensor fields on $S$, respectively, defined
by
\Beqr
&&<h_\orth,f>:=\int_S d^3x f\left[{2\kappa^2\over\sqrt{|q|}}
\left(q(p,p)-{1\over2}(\Tr p)^2\right)-{\sqrt{|q|}\over2\kappa^2}{}^3\!R
\right],\label{GR:HamiltonianConstraint}\\
&&<h_D,X>:=\int_S d^3x (-2X^jD_k p^k_j)=<p,\Lie_X
q>,\label{GR:DiffeomorphismConstraint}\\
&&<p,v>:=\int_S d^3x v_{jk}p^{jk}.
\Eeqr

Let $\T^p_q(S)$ be the set of all smooth $(p,q)$-type tensor field on
$S$, and $\hat q$ be a reference Riemannian metric on $S$. Then by
taking the completion with respect to the inner product
\Beq
(\alpha,\beta)=\sum_{l=0}^{n}\int_S d^3x\, \sqrt{|\hat q|}
\hat q(\hat D^l\alpha,\hat D^l\beta),
\Eeq
we obtain the Sobolev space $H^l_{R}(\T^p_q(S))$ which is a real
Hilbert space. In particular, if we define the space of 3-metrics on
$S$ by
\Beq
\M(S):=\left\{q \in H^2_{R}(\T^0_2(S)) |
\hbox{$q$ is positive definite on $S$} \right\},
\Eeq
$\M(S)$ becomes an open subset of $H^2_{R}(\T^0_2(S))$.  Hence it has
a natural Hilbert manifold structure, and its tangent space and
cotangent space are both isomorphic to the original Hilbert space:
\Beq
T_q(\M(S))\cong T_q^*(\M(S))\cong H^2_{R}(\T^0_2(S)).
\Eeq
Further, since the operator $L$ defined by
\Beq
(\alpha,\beta)=<\alpha,L\hat\beta>,
\Eeq
where $\hat\beta^{jk}=\hat q^{jl}\hat q^{km}\beta_{lm}$, is given by
$L=\sum_{n=0}^2(-1)^n\hat\Delta^n$ and elliptical, it defines an
injection from $T_q^*(\M(S))$ into the space of linear functionals on
$H^0_{R}(\T^0_2(S))$.  Under the identification by this mapping the
momentum $p$ can be regarded as an element of $T_q^*(\M(S))$.

 From this observation the action \Eq{GR:CanonicalAction} defines a
canonical system $(\Gamma,\omega,H)$ with an infinite number of
constraints:
\Beqr
&& <h_D, X>=0 \quad \forall X\in\vfield(S),\\
&& <h_\orth,f>=0 \quad\forall f\in\F(S),
\Eeqr
where the total phase space $\Gamma$ is given by the cotangent bundle
$T^*\M(S)$, the symplectic form formally by $\omega=<\delta
p\wedge\delta q>$, and the hamiltonian $H$ by
\Beq
H=<h_\orth,N>+<h_D,\nu>.
\Eeq
Though we can give an exact expression for $\omega$ by introducing the
basis of $H^2_{R}(\T^0_2(S))$, we will not do it because the
argument in this section is formal.  Since the hamiltonian is
written as a linear combination of the constraint functionals, this
canonical system is a totally constrained system.

\subsection{Dynamical foliation of the phase space and $\DiffG{M}$-classes}

Let $\Sigma_D$ and $\Sigma_H$ be the submanifolds of $\Gamma$ defined by
\Beqr
&&\Sigma_D:=\left\{u\in\Gamma | <h_D,X>(u)=0 \forall X\in\vfield(S)\right\},\\
&&\Sigma_H:=\left\{u\in\Gamma | <h_\orth,f>(u)=0 \forall f\in\F(S)\right\}.
\Eeqr
Then from the Poisson bracket structure among the constraints,
\Beqr
&&\left\{<h_D,X_1>,<h_D,X_2>\right\}=<h_D,[X_1,X_2]>,
\label{GR:ConstraintAlgebra1}\\
&&\left\{<h_D,X>,<h_\orth,f>\right\}=<h_\orth,\Lie_X f>,
\label{GR:ConstraintAlgebra2}\\
&&\left\{<h_\orth,f_1>,<h_\orth,f_2>\right\}=<h_D,f_1Df_2-f_2Df_1>,
\label{GR:ConstraintAlgebra3}
\Eeqr
the infinitesimal canonical transformation $X_H$ generated by $H$ is
tangential both to $\Sigma_D$ and $\Sigma_H$, and each integration
curve of $X_H$ on $\Sigma_D\cap\Sigma_H$ yields a solution to the
Einstein equations. However, this correspondence is not one-to-one
because the same spacetime allows an infinite number of different
slicings with the same $N$ and $\nu$. Further for a different choice
of the lapse function $N$ and the shift vector $\nu$ yields a
different curve in $\Gamma$ for the same spacetime.

This ill correspondence between the spacetime solutions to the
Einstein equations and the curves in the phase space, which arises due
to the general covariance of general relativity, can be made
well-defined by considering the subspace spanned by the integration
curves in stead of each curve.  To see this, let us denote the set of
constraints symbolically by $h_\alpha$ and the corresponding
infinitesimal canonical transformations by $Y_\alpha=X_{h_\alpha}$.
Then from the first class nature of $h_\alpha$ shown above,
$\{Y_\alpha\}$ yields an involutive system on $\Sigma_D\cap\Sigma_H$:
\Beq
[Y_\alpha,Y_\beta]\approx -\sum_\gamma c^\gamma_{\alpha\beta}Y_\gamma,
\Eeq
where $c^\alpha_{\beta\gamma}$ is a set of functions on $\Gamma$.
Hence we obtain a foliation of the constraint submanifold
$\Sigma_D\cap\Sigma_H=\cup_\lambda C_\lambda$ where each leaf
$C_\lambda$ is a connected component of the integration submanifolds.

For an arbitrary non-degenerate curve $\gamma$ contained in a leaf
$C_\lambda$, its tangent vector $X$ is written in terms of some
set of functions $N^\alpha$ as $X=\sum_\alpha N^\alpha Y_\alpha$
because $\{Y_\alpha\}$ spans the tangent space of $C_\lambda$ at
any point.  Hence it is an integration curve of the hamiltonian flow
for the hamiltonian $H=\sum_\alpha N^\alpha h_\alpha$,
and corresponds to some spacetime solution $(M,g)$ to the Einstein equations.
Further, if two curves $\gamma_1$ and $\gamma_2$ are contained in
the same leaf and intersect with each other at a point $u$, they
correspond to solutions to the Einstein equations with the same
initial data for different lapse functions and shift vectors. Hence
from the uniqueness of the initial value problem for the Einstein equations
the corresponding spacetime solutions $(M,g_1)$ and $(M,g_2)$ are
isometric if they are maximally extended.  The same conclusion holds
even if these two curves do not intersect. For  one can find another
curve $\gamma_3$ in the same leaf which intersects both with $\gamma_1$
and $\gamma_2$, which implies that $(M,g_3)\cong (M,g_1)$ and $(M,g_3)
\cong (M,g_2)$, hence $(M,g_1)\cong(M,g_2)$. Therefore all the curves
contained in the same leaf corresponds to a unique 4-dimensional
diffeomorphism class of the spacetime solutions to the Einstein
equations.

We can further show that this correspondence is one-to-one.  Take two
curves $\gamma_1\subset C_1$ and $\gamma_2\subset C_2$ and suppose
that the corresponding spacetime solutions $(M,g_1)$ and $(M,g_2)$ are
isometric. Then there exist isometric diffeomorphisms to a spacetime
$(\tilde M,\tilde g)$, $\Phi_1:(M,g_1)\maps(\tilde M,\tilde g)$ and
$\Phi_2:(M,g_2)\maps(\tilde M,\tilde g)$. Let $S_1 =
\Phi_1(\{t_1\}\times S)$ and $S_2=\Phi_2(\{t_2\}\times S)$ be two
space-like constant-time hypersurfaces in $\tilde M$, and choose two
families of time slicings in $\tilde M$ such that they contain a
common time slice, one of them contains $S_1$ and the other
$S_2$. Further let the two curves in the phase space determined by
these slicings be $\gamma_3$ and $\gamma_4$.  Then from the
construction $\gamma_3\cap\gamma_1\not=\emptyset$,
$\gamma_4\cap\gamma_2\not=\emptyset$ and
$\gamma_3\cap\gamma_4\not=\emptyset$.  This implies that there is a
curve which connects a point in $C_1$ and a point $C_2$. Hence from
the connectedness of each leaf $C_1$ and $C_2$ must coincide with each
other.

Thus we have found that the connected components of the integration
manifolds of the involutive system $\{Y_\alpha\}$ are in one-to-one
correspondence with the 4-dimensional diffeomorphism classes of the
spacetime solutions to the Einstein equations. In other words the
classical dynamics of general relativity is completely determined by
the foliation of the constraint submanifold in terms of the
infinitesimal canonical transformations generated by the constraints.
We can discard the lapse function and the shift vector, or the
corresponding hamiltonian.  For this reason we can simply say that the
canonical theory of general relativity is given by a totally
constrained system $(T^*\M(S),\omega,\{h_D,h_\orth\})$.
We will call each leaf of the foliation \newterm{a causal submanifold}.

\subsection{Elimination of the diffeomorphism constraints}

As the structure constants of the Poisson brackets among $h_D$ are
genuinely constant from \Eq{GR:ConstraintAlgebra1}, the corresponding
infinitesimal canonical transformations are involutive on the whole
phase space and generate the action of the diffeomorphism group of
$S$, $\DiffG{S}$ where the subscript $0$ denotes the connected
component containing the unit element. Since all the measurable
quantities are invariant under these transformations, it is desirable
to eliminate this kinematical gauge symmetry from the canonical
theory, especially when one consider the quantization of the
theory. Now we will show that the classical dynamics has the same
structure as above even after the elimination of this gauge freedom.

First of all note that for a $\F(S)$-valued functional $\tilde N$ on
$\Gamma$ which transforms covariantly under $\DiffG{S}$ as
\Beq
\tilde N(a_* u)=a_*(\tilde N(u)) \quad \forall a\in\DiffG{S}, \quad
\forall u\in\Gamma,
\Eeq
$<h_\orth,\tilde N>$ is invariant under $\DiffG{S}$ as a
functional on $\Gamma$ from
\Beq
\left\{<h_\orth,\tilde N>,<h_D,X>\right\}
=<\Lie_X h_\orth,\tilde N>+<h_\orth,\Lie_X \tilde N>=0.
\Eeq
Similarly for a $\vfield(S)$-valued functional $\tilde \nu$ on
$\Gamma$ which is $\DiffG{S}$-covariant, $<h_D,\tilde \nu>$ is
invariant under $\DiffG{S}$.  Further by inspecting the argument on
the correspondence between a curve in a causal submanifold and the
hamiltonian flow generated by the hamiltonian
$H=<h_\orth,N>+<h_D,\nu>$ one easily sees that $N$ and $\nu$ can be
replaced by some appropriate $\DiffG{S}$-covariant functionals $\tilde
N$ and $\tilde\nu$. Hence the connected integration surfaces of the
involutive system generated by the $\DiffG{S}$-invariant functionals
$<h_\orth,\tilde N>$ and $<h_D,\tilde \nu>$ give the same foliation as
that given by $<h_D,\nu>$ and $<h_\orth,N>$.

Further if $<h_D,\xi>\not=0$ at a point $u\in\Gamma$ for some
$\xi\in\vfield(S)$, there exists a functional $\tilde
\xi:\Gamma\maps\vfield(S)$
such that $<h_D,\tilde \xi>\not=0$ at the same point $u$. Thus
$\Sigma_D$ can be redefined as
\Beq
\Sigma_D=\SetDef{u\in\Gamma}{<h_D,\tilde\nu>(u)=0
\forall \tilde\nu:\Gamma\maps\vfield(S);\DiffG{S}\r{-covariant}}.
\Eeq
Similarly $\Sigma_H$ can be expressed as
\Beq
\Sigma_H=\SetDef{u\in\Gamma}{<h_\orth,\tilde N>(u)=0
\forall \tilde N:\Gamma\maps\F(S);\DiffG{S}\r{-covariant}}.
\Eeq

These arguments indicate that the original canonical system can be
naturally projected on $\Gamma/\DiffG{S}$. To confirm this, let us
denote all the functions on $\Gamma$ which are invariant under
$\DiffG{S}$ by $\F_{\r{inv}}$:
\Beq
\F_{\r{inv}}:=\SetDef{f\in\F(\Gamma)}{\left\{f,<h_D,\xi>\right\}=0
\forall \xi\in\vfield(S)}.
\Eeq
Then it is easily shown that $\F_{\r{inv}}$ is closed with respect to
the Poisson algebra and $\F_{\r{inv}}\F_D=\F_D$ where
\Beq
\F_D:=\SetDef{f\in\F_{\r{inv}}}{f|_{\Sigma_D}=0}.
\Eeq
Further since $\{f,<h_D,\xi>\}=0$ implies that $X_f$ is tangential to
$\Sigma_D$, $\{f,g\}=-X_f g$ vanishes on $\Sigma_D$ for
$f\in\F_{\r{inv}}$ and $g\in\F_D$. Hence $\{\F_{\r{inv}},\F_D\}=\F_D$.
This implies that $\F_D$ forms an ideal of $\F_{\r{inv}}$ and
the Poisson bracket in $\F_{\r{inv}}$ naturally induces a Poisson
bracket in $\A_{\r{inv}}:=\F_{\r{inv}}/\F_D$. Each element of
$\A_{\r{inv}}$ is just a set of functions in $\F_{\r{inv}}$ which
coincide with each other on $\Sigma_D$.

Let $\pi:\Gamma\maps\Gamma/\DiffG{S}$ be the natural projection and
put $\Gamma_{\r{inv}}:=\pi(\Sigma_D)$. Then from the arguments above
$\Gamma_{\r{inv}}$ is characterized as
\Beq
\Gamma_{\r{inv}}=\SetDef{u\in\Gamma/\DiffG{S}}{<h_D,\tilde \nu>(u)=0
\forall\tilde\nu:\Gamma\maps\vfield(S);\DiffG{S}\r{-covariant}},
\Eeq
and $\A_\r{inv}$ is naturally identified with $\F(\Gamma_\r{inv})$.
Further the constraint $h_D$ is trivialized on $\Gamma_\r{inv}$ and the
causal submanifolds in $\Sigma_D\cap\Sigma_D$ is bijectively mapped to the
causal submanifolds in $\pi(\Sigma_H)\cap\Gamma_\r{inv}$ determined
from $<h_\orth,\tilde N>|_{\Sigma_D} \in \Gamma_\r{inv}$.

Let $h_\alpha\in\A_\r{inv}$ be a generating set of all the functions
of the form $<h_\orth,\tilde N>|_{\Sigma_D}$ such that
\begin{itemize}
\item[i)] For any $\DiffG{S}$-covariant functional
$\tilde N:\Gamma\maps\F(S)$ there exists a set of elements
$\lambda^\alpha\in\A_\r{inv}$ such that $<h_\orth,\tilde N>
=\sum_\alpha \lambda^\alpha h_\alpha$,
\item[ii)] $\pi(\Sigma_H)\cap\Gamma_\r{inv}=\SetDef{u\in\Gamma_\r{inv}}
{h_\alpha(u)=0 \forall \alpha}$,
\item[iii)] $\{h_\alpha,h_\beta\}=\sum_\gamma c^\gamma_{\alpha\beta}h_\gamma$.
\end{itemize}
Further let us denote $\pi(\Sigma_H)\cap\Gamma_\r{inv}$ by the same
symbol $\Sigma_H$.  Then the arguments so far shows that the canonical
dynamics of general relativity is described by the totally constrained
system $(\Gamma_\r{inv}, \omega_\r{inv},\{h_\alpha\})$ and the causal
submanifolds in $\Sigma_H$ is one-to-one correspondence with the
$\DiffG{M}$-class of the spacetime solutions to the Einstein
equations.

\subsection{Cotangent-bundle structure of $\Gamma_\r{inv}$}

In the last statement of the previous subsection $\omega_\r{inv}$ is
understood to be the symplectic form corresponding to the Poisson
brackets in $\A_\r{inv}$. Hence in order to make the statement
rigorous it should be shown that $\Gamma_\r{inv}$ has a manifold
structure and the required symplectic form exits. Now we prove these
facts by showing that $\Gamma_\r{inv}$ can be identified with
$T^*(\M(S)/\DiffG{S})$ and $\omega_\r{inv}$ coincides with the
canonical symplectic form corresponding to the cotangent bundle.

\begin{figure}
$$
\begin{array}{cccc}
T^*(\M(S)) & \supset\Sigma_D & \mapsnamed{\phi_1} & \M(S)\\
	   & \pi \downarrow  &                        &\pi\downarrow\\
T^*(\M(S))/\DiffG{S}
&\supset\Gamma_\r{inv}&\mapsnamed{\phi_2}&\M(S)/\DiffG{S}\\
           & \semapsnamed{j} &             &\nemapsnamed{\phi_3}\\
	   & 		     &T^*(\M(S)/\DiffG{S})&
\end{array}
$$
\caption{\label{Diagram:Superspace}}{}
\end{figure}

Let $\phi_1:\Sigma_D\maps \M(S)$ be the restriction of the natural projection
from $T^*(\M(S))$ to $\M(S)$. Then, since the diffeomorphism constraint
implies that $p$ vanishes on the subspace of $T_*(\M(S))$ spanned by
the tangent vectors to the $\DiffG{S}$-orbits from
\Eq{GR:DiffeomorphismConstraint}, $\phi_1$ is surjective and induces
a surjective mapping $\phi_2:\Gamma_\r{inv}\maps\M(S)/\DiffG{S}$ such
that $\pi\phi_1=\phi_2\pi$(See the diagram in
Fig.\ref{Diagram:Superspace}).  Let $\eta_q\in T_q(\M(S))$ be a vector
tangent to a $\DiffG{S}$-orbit passing through $q\in\M(S)$.  Then,
since it is written in terms of a vector field $X\in\vfield(S)$ as
$\eta_q=\Lie_Xq$, it follows from \Eq{GR:DiffeomorphismConstraint}
that for $(p,q)\in \Sigma_D$,
\Beq
(p,q)(\eta_q)=<p,\eta_q>=<p,\Lie_X q>=0.
\Eeq
Further for $a\in\DiffG{S}$ and $v\in T_q(\M(S))$
\Beq
a_*(p,q)(a_*v)=<a_*p,a_*v>=<p,v>=(p,q)(v),
\Eeq
from the diffeomorphism invariance of $<p,v>$. Hence there is an
injection $j:\Gamma_\r{inv}\maps T^*(\M(S)/\DiffG{S})$ such that
for $v\in T_q(\M(S))$
\Beq
j\pi(p,q)(\pi v)=<p,v>.
\label{Def:j}\Eeq
It is easily checked that for the natural projection
$\phi_3:T^*(\M(S)/\DiffG{S})\maps \M(S)/\DiffG{S}$, $\phi_2=\phi_3j$
holds and that $j$ is surjective.  Thus $\Gamma_\r{inv}$ can be
identified with $T^*(\M(S)/\DiffG{S})$.

Next let us show that the symplectic form $\omega_\r{inv}$ induced from
the cotangent bundle structure of $\Gamma_\r{inv}$ is equivalent to
the Poisson brackets in $\A_\r{inv}\cong\F(\Gamma_\r{inv})$ derived from
the symplectic form $\omega$ in $\Gamma$. From now on we identify
$\Gamma_\r{inv}$ with $T^*(\M(S)/\DiffG{S})$ and write $j\pi$ simply
as $\pi$.

We first show that each element of $\A_\r{inv}$ uniquely determines a
vector field on $T^*(\M(S)/\DiffG{S})$, which will turn out to be an
infinitesimal canonical transformation on $\Gamma_\r{inv}$. Let
$\vfield_\r{inv}$ be
\Beq
\vfield_\r{inv}=\SetDef{X\in\vfield(\Sigma_D)}{a_*X=X \;
\forall a\in \DiffG{S}},
\Eeq
and for $[f]\in\A_\r{inv}$, let $\vfield_{[f]}$ be
\Beq
\vfield_{[f]}=\SetDef{X\in\vfield_\r{inv}}{\omega(X,Z)=-Zf \;
\forall Z\in\vfield_\r{inv}}.
\label{GR:Xf:def}
\Eeq
Then for each $X\in\vfield_\r{inv}$ $\pi_* X$ obviously defines a
unique vector field on $T^*(\M(S)/\DiffG{S})$. Further let $\theta$
and $\theta_\r{inv}$ be the canonical 1-forms on $T^*(\M(S))$ and
$T^*(\M(S)/\DiffG{S})$, respectively.  Then from the commutativity of
the diagram in Fig.\ref{Diagram:Superspace} and \Eq{Def:j} it follows
that $(\theta_\r{inv})_{\pi(p,q)}(\pi_* X)=\theta_{(p,q)}(X)$ for any
$X\in\vfield_\r{inv}$, i.e.,
\Beq
\theta(X)=\pi^*\theta_\r{inv}(X) \quad \forall X\in\vfield_\r{inv}.
\Eeq
Hence for $\Delta X=X_1-X_2$($X_1,X_2\in\vfield_{[f]}$) and
$Z\in\vfield_\r{inv}$, from
\Beq
0=\omega(\Delta X,Z)=d\theta(\Delta X,Z)=\Delta X(\theta(Z))
-Z(\theta(\Delta X))-\theta([\Delta X,Z]),
\Eeq
it follows that
\Beqr
&0&=\pi_*(\Delta X)(\theta_\r{inv}(\pi_*Z))-\pi_*Z(\theta_\r{inv}(\pi_*X))
-\theta_\r{inv}(\pi_*[\Delta X,Z]) \nonumber\\
&&=d\theta_\r{inv}(\pi_*\Delta X,\pi_*Z)=\omega_\r{inv}(\pi_*\Delta X,\pi_*Z).
\Eeqr
As $\pi_*Z$ can be any vector field on $T^*(\M(S)/\DiffG{S})$, this equation
implies that $\pi_*\Delta X=0$, i.e., $\pi_*X_1=\pi_*X_2$.  Hence
$[f]\in\A_\r{inv}$ determines a unique vector field on $T^*(\M(S)/\DiffG{S})$.
We denote this vector field by $X_{[f]}$.

Next we show that for $X\in\vfield_{[f]}$ and $Y\in\vfield_{[g]}$
$\omega(X,Y)=-\{f,g\}$.  For $X_1,X_2\in\vfield_{[f]}$ and $Y_1,Y_2
\in\vfield_{[g]}$ from the definition (\ref{GR:Xf:def}) it follows that
\Beq
\omega(X_2,Y_2)-\omega(X_1,Y_1)=\omega(X_2-X_1,Y_2)+\omega(X_1,Y_2-Y_1)=0.
\Eeq
Hence $\omega(X,Y)$ depends only on $[f]$ and $[g]$. Obviously
$X_f\in\vfield_{[f]}$ and $X_g\in\vfield_{[g]}$. Therefore
\Beq
\{f,g\}=-\omega(X_f,X_g)=-\omega(X,Y).
\Eeq

With the help of the equations derived so far for
$X\in\vfield_{[f]}$ and $Y\in\vfield_{[g]}$ we obtain
\Beqr
&[\{f,g\}]&=-[\omega(X,Y)]=-[d\theta(X,Y)] \nonumber\\
&&=[Y(\theta(X))-Y(\theta(X))-\theta([X,Y])] \nonumber\\
&&=\pi_*Y(\theta_\r{inv}(\pi_*X))-\pi_*Y(\theta_\r{inv}(\pi_*X))
-\theta_\r{inv}(\pi_*[X,Y]) \nonumber\\
&&=-d\theta_\r{inv}(\pi_*X,\pi_*Y)=-\omega_\r{inv}(X_{[f]},Y_{[g]})
\nonumber\\
&&=\{[f],[g]\}_\r{inv}
\Eeqr
This shows that the Poisson brackets induced from $\omega$
coincides with that defined by $\omega_\r{inv}$.

Note here that the arguments so far are not mathematically rigorous
because $\M(S)/\DiffG{S}$ has conical singularities at metrics with
Killing vectors\CITE{Fischer.A&Marsden1979B}.
Though these singularities may have physical importance in
quantization,  we will not go into this problem in this paper.

We can go further and eliminate all the hamiltonian constraints to
get the fully reduced phase space with a symplectic structure which
represents the true physical degrees of freedom as done by Fischer
and Marsden.  However, we shall not follow this line because we will
then lose the dynamics.

\section{General Totally Constrained Systems}

Now we discuss the dynamics of a generic totally constrained system.
Here a totally constrained system is defined as a triplet of a
phase space, a symplectic form and a set of constraint functions,
$(\Gamma,\omega,\{h_\alpha\})$. For a technical reason we assume that
the phase space is $2n$-dimensional smooth manifold with finite $n$.
Further we assume that the constraints are of first class with the
Poisson brackets given by
\Beq
\{h_\alpha, h_\beta\}=\sum_\gamma c^\gamma_{\alpha\beta}h_\gamma,
\Eeq
where $c^\gamma_{\alpha\beta}$ are functions on $\Gamma$.

On the basis of the arguments in the previous section we understand
that the physical evolution of the system is one-to-one correspondence
with each leaf of the foliation determined by the involutive system
of the infinitesimal canonical transformations $Y_\alpha=X_{h_\alpha}$
on the constraint submanifold $\Sigma_H=\SetDef{u\in\Gamma}{h_\alpha(u)=0
\;\forall\alpha}$. We call each leaf a causal submanifold as so far.
As is clear from the arguments in the previous section, this interpretation
is equivalent to regard that two solutions to the canonical equation of motion
for the hamiltonian $H=\sum_\alpha \lambda^\alpha h_\alpha$ with
arbitrary functions $\lambda^\alpha$ represent the same physical
evolution if they intersect with each other in $\Gamma$.

This is a natural generalization of the argument on the dynamics
of a single totally constrained system with one constraint in \S3.
Now we extend this generalization to the statistical dynamics of
an ensemble.

\subsection{Relative distribution function}

 From this interpretation of dynamics of a single system and the argument
in \S3 it is natural to introduce the relative distribution function
$\rho$ on $\Gamma$ to describe an ensemble, which vanishes outside the
constraint submanifold and is constant on each causal submanifold:
\Beqr
&&Y_\alpha \rho=0 \quad \forall\alpha, \label{Multi:EqForRho1}\\
&&h_\alpha \rho=0 \quad \forall\alpha. \label{Multi:EqForRho2}
\Eeqr
Let us define an acausal submanifold as a submanifold of $\Gamma$ which
intersects with causal submanifolds transversally. Then for any acausal
submanifold $\Sigma$ and for any distribution $\rho|_\Sigma$ on $\Sigma$
a solution to these equations which coincides with $\rho|_\Sigma$
on $\Sigma$ is unique, if it exists, on the \newterm{causal development} of
$\Sigma$
defined by
\Beq
D(\Sigma):=\bigcup_{C\cap\Sigma\not=\emptyset}C
\label{Multi:CausalDevelopment:Def}\Eeq
where $C$ runs over causal submanifolds.
However, such solution may not exists in general. In fact
the following theorem holds.

\begin{theorem}\label{Multi:ConsistencyCondition}
In order that there exists a solution to
\Eqs{Multi:EqForRho1}{Multi:EqForRho2} for arbitrary initial data
on any acausal submanifold, the following condition should be
satisfied:
$$
\sum_\gamma\{h_\gamma,c^\gamma_{\alpha\beta}\}\approx0.
$$
This condition is satisfied if and only if there exists a function
$f\not=0$ such that for $c'{}^\gamma_{\alpha\beta}$ corresponding
to the constraints $h'_\alpha=f h_\alpha$
$$
c'_\alpha:=\sum_\beta c'{}^\beta_{\alpha\beta}\approx0
$$
holds.
\end{theorem}
\begin{proof}
Since $\rho$ is a distribution, to be exact,
\Eqs{Multi:EqForRho1}{Multi:EqForRho2} are expressed as
\begin{eqnarray*}
&&<h_\alpha\rho,\phi>:=<\rho,h_\alpha\phi>=0,\\
&&<Y_\alpha\rho,\phi>:=<\rho,-Y_\alpha\phi>=0,
\end{eqnarray*}
where $\phi$ is an arbitrary smooth function with a compact support
on $\Gamma$. However, since the commutators among $Y_\alpha$'s are
given by
$$
[Y_\alpha,Y_\beta]\approx -\sum_\gamma c^\gamma_{\alpha\beta}Y_\gamma,
$$
the consistency condition yields
$$
0=<\rho,[Y_\alpha,Y_\beta]\phi>=<\rho,\sum_\gamma (Y_\gamma
c^\gamma_{\alpha\beta})\phi>.
$$
Hence, noting the relation $Y_\gamma c^\gamma_{\alpha\beta}
=-\{h_\gamma,c^\gamma_{\alpha\beta}\}$, we obtain the first condition
in the theorem.

In order to show the latter half of the theorem, first note that
the Jacobi identity for the Poisson brackets among $h_\alpha$ yields
$$
Y_\alpha c_\beta -Y_\beta c_\alpha+\sum_\gamma c^\gamma_{\alpha\beta}
\approx\sum_\gamma\{h_\gamma,c^\gamma_{\alpha\beta}\},
$$
where $c_\alpha:=\sum_\beta c{}^\beta_{\alpha\beta}$.  From this
it immediately follows that the first condition of the theorem holds
if $c_\alpha\approx0$.

On the other hand for $h'_\alpha=fh_\alpha$ $c_\alpha$ changes as
$$
c'_\alpha=c_\alpha-(m-1)Y_\alpha \ln f,
$$
where $m$ is the number of the constraints. Hence the second condition
of the theorem is satisfied if $f$ is a solution to the equation
$$
Y_\alpha f^{m-1}\approx c_\alpha f^{m-1}.
$$
However, if the first condition of the theorem is satisfied, we obtain
$$
Y_\alpha c_\beta -Y_\beta c_\alpha\approx
-\sum_\gamma c^\gamma_{\alpha\beta}.
$$
This is nothing but the consistency condition for the first-order
differential equation system for $f^{m-1}$ above. Hence the first
condition of the theorem implies the second. \QED
\end{proof}

Note that for a matrix function $\Lambda=(\Lambda_\alpha^\beta)$ on
$\Gamma$ with $\det\Lambda\not=0$, the totally constrained system with
the constraints $h'_\alpha=\sum_\beta\Lambda_\alpha^\beta h_\beta$ is
equivalent to the original system. Hence the precise meaning of the
requirement of the theorem is that $\sum_\gamma\{h_\gamma,c^\gamma_{\alpha
\beta}\}$ can be put to zero by
such a transformation and that \Eqs{Multi:EqForRho1}{Multi:EqForRho2}
are consistent only for such choice of the constraints.
This result is interesting in relation to the quantization of the
totally constrained system because this condition implies that the
operators corresponding to $c^\gamma_{\alpha\beta}$ and $h_\alpha$
should commute in a weak sense.

On the basis of this theorem we assume that $c_\alpha=0$ from now on.
Under this condition if we put
\Beq
\rho=\rho_0\prod_\alpha \delta(h_\alpha),
\Eeq
\Eq{Multi:EqForRho1} is automatically satisfied and \Eq{Multi:EqForRho2}
reduces to the equation for the function $\rho_0$,
\Beq
Y_\alpha \rho_0\approx0,
\Eeq
because
\Beqr
&\Lie_{Y_\alpha}\rho&=Y_\alpha \prod_\beta \delta(h_\beta)
+\sum_{\beta,\gamma}c^\gamma_{\alpha\beta}h_\gamma \delta'(h_\beta)
\prod_{\mu\not=\beta}\delta(h_\mu) \nonumber\\
&&=(Y_\alpha \rho_0 - c_\alpha\rho_0)\prod_\mu\delta(h_\mu).
\Eeqr

\subsection{Statistical dynamics in terms of conservative measure on
acausal submanifolds}

Now we show that we can formulate the statistical dynamics for an
ensemble of the totally constrained system without reducing it
to a system without constraint. The basic idea is the same as
that used in \S2 and \S3 for totally constrained systems
with a single constraint.

First note that for any pair of acausal submanifolds $\Sigma_1$ and $\Sigma_2$
the foliation of the constraint submanifold by the constraints uniquely
determines the causal mapping
\Beq
\theta: \Sigma_1\cap D(\Sigma_2)\cap \Sigma_H \maps
\Sigma_2\cap D(\Sigma_1)\cap \Sigma_H,
\Eeq
where $D(\Sigma_1)$ and $D(\Sigma_2)$ are causal developments of
$\Sigma_1$ and $\Sigma_2$, respectively. We extend this causal mapping
to a neighborhood of $\Sigma_H$ by considering a foliation of the
tubular neighborhood such that the intersection of each leaf with $\Sigma_H$
coincides with the foliation of $\Sigma_H$ by $Y_\alpha$. If a measure
$\mu_{\Sigma_0}$ on an acausal submanifold $\Sigma_0$ with its support
contained in $\Sigma_0\cap\Sigma_H$ is given, this causal mapping
uniquely determines a measure $\mu_{\Sigma}$ on $\Sigma\cap D(\Sigma_0)$
with its support contained in $\Sigma_H$ such that for any constant of
motion, i.e., a function $f\in\F(\Gamma)$ which is constant along each
leaf,
\Beq
\int_{\Sigma} f d\mu_{\Sigma} = \int_{\Sigma_0} f d\mu_{\Sigma_0},
\Eeq
if $\support{f}\cap\Sigma_0\subset \domain{\theta}$.  As in the single
constraint systems, this measure can be expressed locally in terms of
$\rho$, $Y_\alpha$ and $\Omega$.

\begin{theorem}\label{Multi:ConservativeMeasure1}
If $c_\alpha=0$, for any distribution $\rho$ satisfying
\Eqs{Multi:EqForRho1}{Multi:EqForRho2} the measure
$$
d\mu_\Sigma=\rho|I_{Y_1}\cdots I_{Y_m}\Omega|_{\Sigma}
$$
is conserved by causal mappings where $m$ is the number of independent
constraints.
\end{theorem}
\begin{proof}
Let us denote $I_{Y_1}\cdots I_{Y_m}(\rho\Omega)$ simply as $\chi$.
Then from the identities
\begin{eqnarray*}
&& d\circ I_X + I_X\circ d=\Lie_X,\\
&& [\Lie_X, I_Y]=I_{[X,Y]},
\end{eqnarray*}
we obtain
$$
d \chi
=-I_{Y_1}dI_{Y_2}\cdots I_{Y_m}(\rho\Omega)
+\sum_{\alpha=1}^m c^1_{1\alpha}(-1)^\alpha\prod_{\beta\not=\alpha}
I_{Y_\beta}(\rho\Omega).
$$
In a similar way we obtain
\begin{eqnarray*}
&&(-1)^{\gamma-1}I_{Y_1}\cdots dI_{Y_\gamma}\cdots I_{Y_m}(\rho\Omega)
=(-1)^\gamma I_{Y_1}\cdots I_{Y_\gamma}dI_{Y_{\gamma+1}}\cdots(\rho\Omega)\\
&&\quad + (-1)^{\gamma-1}\sum_{\alpha=\gamma+1}^m\left[c^\alpha_{\gamma\alpha}
\prod_{\beta\not=\gamma}I_{Y_\beta}(\rho\Omega)-(-1)^{\alpha-\gamma}
c^\gamma_{\gamma\alpha}\prod_{\beta\not=\alpha}I_{Y_\beta}(\rho\Omega)\right].
\end{eqnarray*}
Since $d(\rho\Omega)=0$ and $\Lie_{Y_\alpha}(\rho\Omega)=0$, summation of
these equations over $\gamma=1,\ldots,m$ yields
$$
d\chi
=\sum_{\gamma=1}^m(-1)^{\gamma-1}c_\gamma\sum_{\beta\not=\gamma}
I_{Y_\beta}(\rho\Omega)=0.
$$
Hence for a pair of acausal submanifolds $\Sigma_1$ and $\Sigma_2$
such that $D(\Sigma_1)=D(\Sigma_2)$, from Stokes' theorem on
$(2n-m+1)$-dimensional submanifold $N$ such that $\partial N=
\Sigma_1\cup\Sigma_2\cup\Sigma'$ and $\Sigma'$ is parallel to the
leaves, we obtain
$$
\int_{\Sigma_2}|\chi|-\int_{\Sigma_1}|\chi|
=\pm\int_{\partial N}\chi \mp \int_{\Sigma'}\chi =\pm\int_Nd\chi=0.
$$
\QED
\end{proof}

In realistic situations each acausal submanifold is specified by
a set of independent $m$ functions $\phi_\alpha$ such that
$Y_\alpha\phi_\beta=\{\phi_\beta,h_\alpha\}$ is a regular matrix
as, say,
\Beq
\phi_1=\cdots=\phi_m=0.
\Eeq
Let us call such a set of functions \newterm{instant functions}.
Then the measure given in the previous theorem is expressed in
terms of these instant functions as follows:

\begin{theorem}\label{Multi:ConservativeMeasure2}
If $\phi_1,\cdots,\phi_m$ are instant functions for an acausal
submanifold $\Sigma$, for any $f\in\F(\Gamma)$ the following equality
holds:
$$
\int_\Sigma f\rho|I_{Y_1}\cdots I_{Y_m}\Omega|
=\int_\Gamma f\rho|\det\{h_\alpha,\phi_\beta\}|
\prod_{\gamma}\delta(\phi_\gamma)|\Omega|.
$$
\end{theorem}
\begin{proof}
{}From
\begin{eqnarray*}
&&d\phi_1\wedge\cdots\wedge d\phi_m\wedge I_{Y_1}\cdots I_{Y_m}\Omega\\
&&\quad ={1\over m!}\sum_{\alpha_1\cdots \alpha_m}
\epsilon^{\alpha_1\cdots \alpha_m}d\phi_{\alpha_1}\wedge\cdots
\wedge d\phi_{\alpha_m}\wedge I_{Y_1}\cdots I_{Y_m}\Omega\\
&&\quad ={(-1)^m\over m!}\sum_{\alpha_1\cdots \alpha_m}
\epsilon^{\alpha_1\cdots \alpha_m}I_{Y_1}(d\phi_{\alpha_1}\wedge\cdots
\wedge d\phi_{\alpha_m}\wedge I_{Y_2}\cdots I_{Y_m}\Omega)\\
&&\qquad -{(-1)^m\over (m-1)!}\sum_{\alpha_1\cdots \alpha_m}
\epsilon^{\alpha_1\cdots \alpha_m}(I_{Y_1}d\phi_{\alpha_1})
d\phi_{\alpha_2}\wedge\cdots
\wedge d\phi_{\alpha_m}\wedge I_{Y_2}\cdots I_{Y_m}\Omega\\
&&\quad ={(-1)^m\over (m-1)!}\sum_{\alpha_1\cdots \alpha_m}
\epsilon^{\alpha_1\cdots \alpha_m}\{h_1,\phi_{\alpha_1}\}
d\phi_{\alpha_2}\wedge\cdots
\wedge d\phi_{\alpha_m}\wedge I_{Y_2}\cdots I_{Y_m}\Omega\\
&&\quad =\cdots\\
&&\quad =(-1)^{m(m+1)/2}\sum_{\alpha_1\cdots \alpha_m}
\epsilon^{\alpha_1\cdots \alpha_m}\{h_1,\phi_{\alpha_1}\}
\cdots\{h_m,\phi_{\alpha_m}\}\Omega,
\end{eqnarray*}
we obtain
$$
\prod_\alpha\delta(\phi_\alpha)\prod_\beta d\phi_\beta
|I_{Y_1}\cdots I_{Y_m}\Omega|
=|\det\{h_\alpha,\phi_\beta\}|\prod_\gamma\delta(\phi_\gamma)|\Omega|.
$$
By multiplying $f\rho$ on the both sides of this equation and integrating
over $\Gamma$, we obtain the equation in the theorem. \QED
\end{proof}

 From these theorems we can formulate the statistical dynamics of an
ensemble of the totally constrained system with multiple constraints
in the following way.  First, from the data set obtained by measurements,
pick up a set of instant functions $\phi_{(1)\alpha}$ which take
a common set of values in the data set. For simplicity assume that
these values are all zero, and let the corresponding acausal submanifold
in $\Gamma$ be $\Sigma_1$, and define the measure $d\nu$ by
\Beq
d\nu:=c |I_{Y_1}\cdots I_{Y_m}\Omega|
\Eeq
where $c$ is some fixed positive constant of motion. Then the other data
uniquely determines the distribution $\rho$ on
$\Sigma_1$ through the formula
\Beq
<f>_{\Sigma_1}=\int_{\Sigma_1}f\rho d\nu.
\Eeq
Extend this distribution $\rho$ over $D(\Sigma_1)$ by the evolution
equations \Eqs{Multi:EqForRho1}{Multi:EqForRho2}. Then for another
set of instant functions $\phi_{(2)\alpha}$ corresponding to an
acausal submanifold $\Sigma_2$, the expectation value of a function
$f\in\F(\Gamma)$ on that submanifold is given by
\Beq
<f>_{\Sigma_2}=\int_{\Sigma_2}f\rho d\nu,
\Eeq
if $\support{f}\in D(\Sigma_1)$. Of course we do not need the explicit
knowledge on the acausal submanifolds, because from
\Theorem{Multi:ConservativeMeasure2} the expectation values are
written as an integration over $\Gamma$ in terms of measures expressed
by the constraints and the instant functions.

Like the case of a single constraint system we can define that a set of
functions $\phi_\alpha$ are good time variables if the natural measure
$\prod_\alpha\delta(\phi_\alpha-\tau_\alpha)|\Omega|$ on a set of
acausal submanifolds
\Beq
\Sigma(\tau_\alpha):=\SetDef{u\in\Gamma}{\phi_\alpha(u)=\tau_\alpha}
\Eeq
coincides with the conserved measure.  This condition is expressed as
\Beq
c|\det\{h_\alpha,\phi_\beta\}|\approx f(\phi_1,\cdots,\phi_m),
\Eeq
where $c$ is some positive constant of motion and $f$ is some function
of $m$ variables.

\section{Discussion}

In this paper we have shown that the dynamics of a classical totally
constrained system can be consistently formulated without reducing it
to an unconstrained system by solving the constraints or referring to
a special time variable.  The basis idea has been to consider the
relative distribution function which is constant on each leaf of the
foliation defined by the infinitesimal canonical transformations
generated by the constraint functions, and to normalize it on an
acausal submanifold which is transversal to the foliation in terms of
the conservative measure.

The fact that we can formulate the classical statistical dynamics of a
totally constrained system without referring a special time variable is
very important for considering a quantum theory of the totally
constrained system because a quantum theory has a similar structure to
the classical statistical dynamics in general.  In fact in the next
paper we will show that by introducing a similar foliation structure
into a state space of quantum theory and by considering a relative
probability amplitude in stead of the relative distribution
we can construct a consistent
formulation of the quantum dynamics of a totally constrained system
without referring to a special time variable under some restrictions.

Though the main purpose of the present paper has been to give a
basic motivation for the quantum formulation developed in the subsequent
papers, the results obtained in the paper may be interesting by themselves.
In particular the fact that the conservative measure can be written only
by the canonical volume form and the constraint functions even for
multi-constrained systems seems to be useful in the arguments of
the probability distribution of the initial condition of the universe
in the classical framework and stochastic treatment of general relativity.

Of course the expression for the measure given in this paper
cannot be applied to general relativity directly because we have
only considered systems with finite degrees of freedom.  However,
it seems possible to extend the formulation to general
relativity by taking an appropriate limit. To examine this limiting
procedure explicitly in some simple situations such as the perturbation
theory of general relativity on cosmological background spacetimes and
spherical black hole spacetimes with scalar fields will be interesting.

\section*{Acknowledgments}

This work started when I participated in the program ``Geometry and
Gravity'' in Newton Institute for Mathematical Science.  The author
would like to thank the participants of the program for valuable
discussions and the staff of the institute for their hospitality.
This work was supported by the Grant-In-Aid
for Scientific Research of the Ministry of Education, Science and
Culture in Japan(05640340).

\end{document}